\def\aa{{A\&A}}
\def\aas{{A\&AS}}
\def\aj{{AJ}}
\def\annrev{{ARA\&A}}
\def\apj{{ApJ}}
\def\apjs{{ApJS}}
\def\baas{{BAAS}}
\def\mnras{{MNRAS}}
\def\nat{{Nature}}
\def\pasp{{PASP}}

\documentclass[cup5b]{caps}
\usepackage{graphicx}
\usepackage{amssymb}
\usepackage{ociwsymp4e}  

\HeadText{F. Calura, et al.}  

\begin{document}

\pagenumbering{arabic}

\author[]{F. Calura$^{1}$, F. Matteucci$^{1}$, M. Dessauges-Zavadsky$^{2}$, S. D'Odorico$^{3}$, 
J. X. Prochaska$^{4}$, G. Vladilo$^{5}$
\\
(1) Dipartimento di Astronomia, Universit\'a di Trieste, Trieste, Italy\\
(2) Observatoire de Geneve, Switzerland\\
(3) European Southern Observatory, Garching bei Muenchen, Germany\\
(4) UCO/Lick Observatory, University of California, Santa Cruz, CA, USA\\
(5) INAF, Osservatorio Astronomico di Trieste, Trieste, Italy}
%
%

\chapter{Chemical Evolution of Damped Lyman Alpha Systems}

\begin{abstract}
By means of detailed  chemical evolution models for galaxies of different morphological types (i.e. spirals, irregular/starburst galaxies 
and ellipticals) we study the nature of Damped Lyman-$\alpha$ systems. Our concern is to infer which systems 
represent likely candidates for the DLA population and which do not. By focusing on individual systems, we can derive some constraints 
on both the nature of the associated galaxy and its age. 
Our results indicate that, owing to their high metallicities and [$\alpha$/Fe] ratios, big spheroids represent 
unlikely DLA candidates whereas spirals (observed at different galactocentric distances) and irregulars are ideal sites where DLA absorptions can occur. 
\end{abstract}

\section{Introduction}
Damped Lyman-$\alpha$ (DLA) systems represent the objects with the highest  
neutral gas content in the high-redshift universe ($N(HI) \ge 2 \cdot 10^{20} cm^{-2}$) and their ISM shows traces of 
substantial chemical 
enrichment (their 
metallicities can span from $\sim 1/100$ up to $\sim 1/3$ of
the solar value, see Prochaska, these proceedings).
For these reasons, DLAs are believed to represent the 
progenitors of the present-day galaxies. 
The very high $N_{HI}$ values allow very accurate abundance measurements   
for many low ionization species (e.g. SiII, FeII, ZnII), thus DLAs 
represent very helpful tools in the study of early galactic chemical evolution.
In this paper, by means of detailed chemical evolution models for galaxies 
of different morphological types (spirals, irregulars, ellipticals)
we aim at reconstructing the star formation history 
(or histories) which gave rise to the abundance patterns observed in DLAs.
In chemical evolution
models absolute abundances usually depend on all the model
assumptions, whereas the abundance ratios depend only on
nucleosynthesis, stellar lifetimes and initial mass function (IMF).
Abundance ratios can therefore
be used as cosmic clocks if they involve two elements formed on quite
different timescales, typical examples being [$\alpha$/Fe] and [N/$\alpha$]
ratios. These ratios, when examined together with [Fe/H], or any other
metallicity tracer such as [Zn/H], allow us to clarify the
particular history of star formation involved, as shown by
Matteucci (2001). 
In a regime of high star formation rate we
expect to observe overabundances of $\alpha$-elements for a large
interval of [Fe/H], whereas the contrary is expected for a regime
of low star formation. This is due to the different roles played by the
supernovae (SNe) of type II relative to the SNe of type Ia. These latter, 
in fact, are believed to be responsible for the bulk of Fe and Fe-peak
element production and occur on much longer timescales than SNe II,
which are responsible for the production of the $\alpha$-elements
(i.e. O, Ne, Mg, Si, Ca). 
For this reason, the analysis of the
relative abundances can enable us to have important hints on
the nature and age of the (proto-) galaxies which give rise to DLA
systems.
\section{The chemical evolution models}
A chemical evolution model (see Matteucci, these proceedings) allows one to follow in detail the evolution of the abundances 
of several chemical
species, starting from the matter reprocessed 
by the stars and restored into the ISM through stellar winds and supernova
explosions.
According to our scheme, elliptical galaxies form as the result of a rapid 
collapse of a homogeneous sphere of
primordial gas (Matteucci 1994) where star formation is taking place at 
the same time 
as the collapse proceeds, and evolve as
"closed-boxes", i.e. without any interaction with the external environment. 
%
The model for the spiral is calibrated on the chemical features of the Milky-Way (MW, Chiappini et al. 2001).
It is assumed to form as a result of two main infall episodes
(Chiappini et al. 1997). During the first episode the halo forms and 
the gas shed by the 
halo rapidly
gathers in the center leading to the formation of the bulge. 
During the second episode, 
a slower infall
of external gas gives rise to the disk with the gas 
accumulating faster in the inner than 
in the outer
region ("inside-out" scenario, Matteucci \& Fran\c cois, 1989). 
Irregular galaxies are assumed to form owing to a continuous infall 
of pristine gas, until a mass of
$\sim 10^{9}M_{\odot}$ is accumulated.
The star formation rate can proceed either in bursts 
separated by
long quiescent periods or at a low regime but continuously.
The nucleosynthesis prescriptions are common to all the models and 
include: the yields 
of Thielemann, Nomoto \& Hashimoto (1996) for massive stars
(M $> 10 M_{\odot}$), 
the yields of van den Hoeck \& Groenewegen (1997) for
low and intermediate mass stars ($0.8 \le M/M_{\odot} \le 8$) and the yields of 
Nomoto et al. (1997) for type Ia SNe. For the evolution of Zn and Ni we adopted
the nucleosynthesis prescriptions of Matteucci et al. (1993), where the yields of both elements are assumed to scale as the Fe ones in both Ia and II supernovae. 
Such an assumption provides a satisfactory fit to the [Zn/Fe] and [Ni/Fe] vs [Fe/H] patterns 
observed in Galactic stars in the solar neighbourhood (Calura et al. 2003).
We assume a Salpeter (1955) IMF for irregular/starburst and elliptical galaxies, and a Scalo (1986) for spirals.

\section{Results}
\subsection{The metallicity evolution}
Zn is a very reliable metallicity indicator since, 
unlike Fe, it is not affected by dust depletion (Savage \& Sembach 1996), hence its measured abundances 
do not need dust corrections.
Figure 1.1 shows the evolution of the [Zn/H] ratio versus redshift for galaxies 
of different morphological types, compared to the
values measured in DLAs by various authors.  
                \begin{figure}
                  \centering
                  \includegraphics[width=10cm,height=10cm,angle=0]{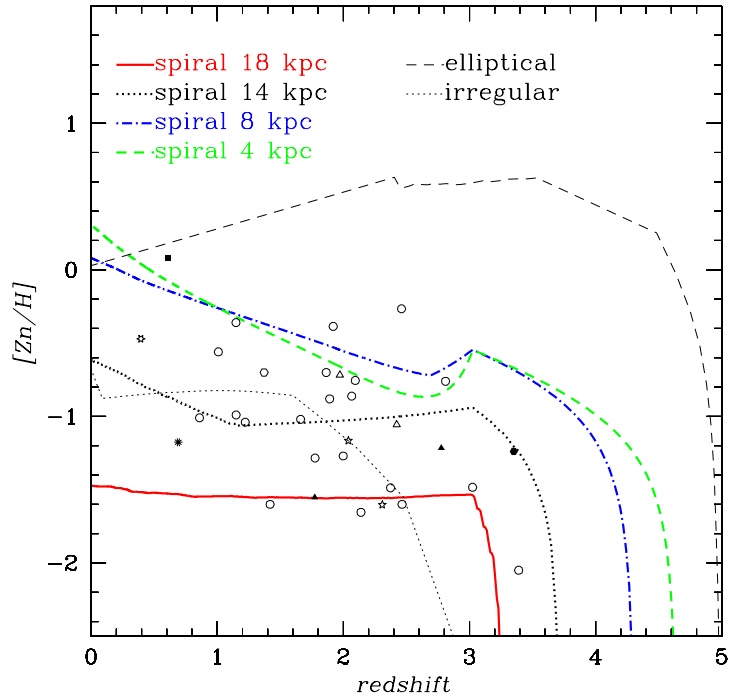}
                  \caption{Redshift evolution of [Zn/H] observed by various authors (see Calura et al. 2003 and references therein) 
and as predicted by chemical evolution models for ellipticals, spirals and irregulars). }
                   \label{sample-figure}
                \end{figure}
We assume a standard $\Lambda$CDM cosmology ($\Omega_{m}=0.3, \Omega_{\Lambda}=0.7$ and $h=0.65$)
and for the sake of simplicity the same formation
redshift for all galaxies ($z_f=5$). Though this represents a rather unrealistic picture, 
since it is unlikely that all DLA systems started forming stars at the same epoch, this figure 
provides clear indications on which systems can represent possible DLA candidates and which cannot. 
In fact, the metallicity levels reached by ellipticals are in general considerably higher than 
the ones observed in DLAs. 
On the other hand, all the points lie in the regions spanned by the spiral models, 
whose [Zn/H] values are in general undersolar in the external regions and become slightly oversolar in 
the most internal ones, and the irregular model, whose low star formation rate produces 
metallicities very similar to the ones of the spiral outskirts.
The conclusion is that the DLA population is heterogeneous, possibly composed by large disk 
galaxies observed at different galactocentric distances, and small irregular systems. 
\subsection{The [$\alpha$/Fe] ratio}
The presence of dust in DLAs (which is indicated by several clues such 
as the reddening of QSOs in presence of DLAs along the line of sight, see Fall
and Pei 1995) represents a complication in chemical evolution studies, 
since its effect is to deplete some chemical elements (e. g. Fe, Si) more than others (O, Zn) 
thus altering the observed abundances. In order to face this effect, the observations need 
reliable corrections for dust depletion.
Figure 1.2 shows the [Si/Fe] distribution versus redshift (upper panels) and [Fe/H] (lower panels),
as predicted by models of galaxies of
different morphological types and compared with uncorrected (left panels) 
DLA data and corrected (right panels) for dust depletion according to Vladilo (2002).
We notice that, without dust correction, the majority of the observed systems shows strong signs of Si 
enhancement, being the data in the [Si/Fe] vs [Fe/H] plot more consistent with an elliptical abundance 
pattern than with a spiral or irregular one. Once we apply the dust corrections to the data, 
the enhancement observed in all the systems is considerably reduced or vanished and all the points 
become fairly consistent with the spiral and irregular curves. This result is confirmed by the study 
of the abundance ratios between elements not affected by dust depletion, e.g. S and Zn, 
as is shown in figure 1.3. For the few systems considered in this set of data (Centurion et al. 2003), 
only one object (DLA0347-383, Levshakov et al. 2001) out of eight is consistent with the prediction for an elliptical galaxy, 
whereas another DLA 
system (DLA0013-004, Centuri\'on et al. 2000) has a [S/Zn] value lower than the predictions of all the models considered here, 
but the majority of the points is reproduced either by models of spirals and irregulars. 
This is another clear indication that, once the dust corrections are taken into account, 
elliptical galaxies (and big spheroids in general) cannot explain the abundance ratios observed in 
DLAs.

                \begin{figure}
                  \centering
                  \includegraphics[width=10cm,height=10cm,angle=0]{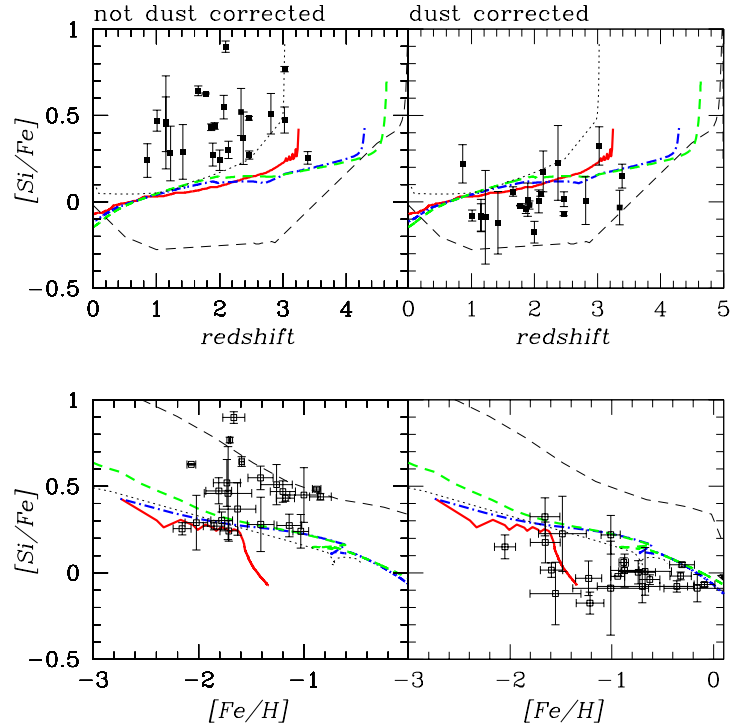}
                  \caption{[Si/Fe] vs redshift (upper panels) and [Fe/H] (lower panels) observed in DLAs (see Calura et al. 2003 and references therein) and compared to chemical evolution models in the case of data not corrected (left panels) and corrected (right panels) for dust depletion. The models are the same as 
in figure 1.1.}
                   \label{sample-figure}
                \end{figure}
 
                \begin{figure}
                  \centering
                  \includegraphics[width=10cm,height=10cm,angle=0]{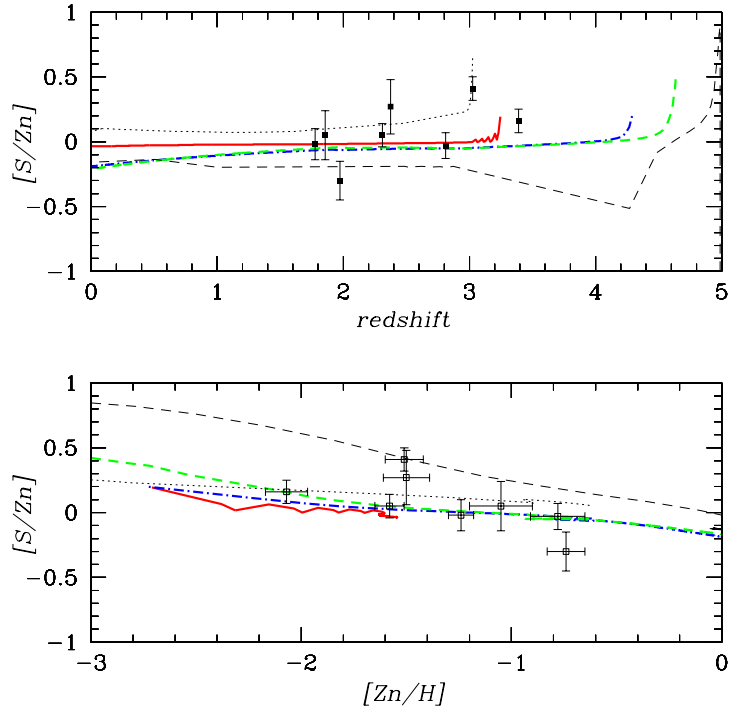}
                  \caption{[S/Zn] vs redshift (upper panel) and [Zn/H] (lower panel) observed in DLAs and compared to chemical evolution models. The models are the same as in figure 1.1.}
                   \label{sample-figure}
                \end{figure}
 
\subsection{The [N/$\alpha$] ratio}
               \begin{figure}
                  \centering
                  \includegraphics[width=10cm,height=10cm,angle=0]{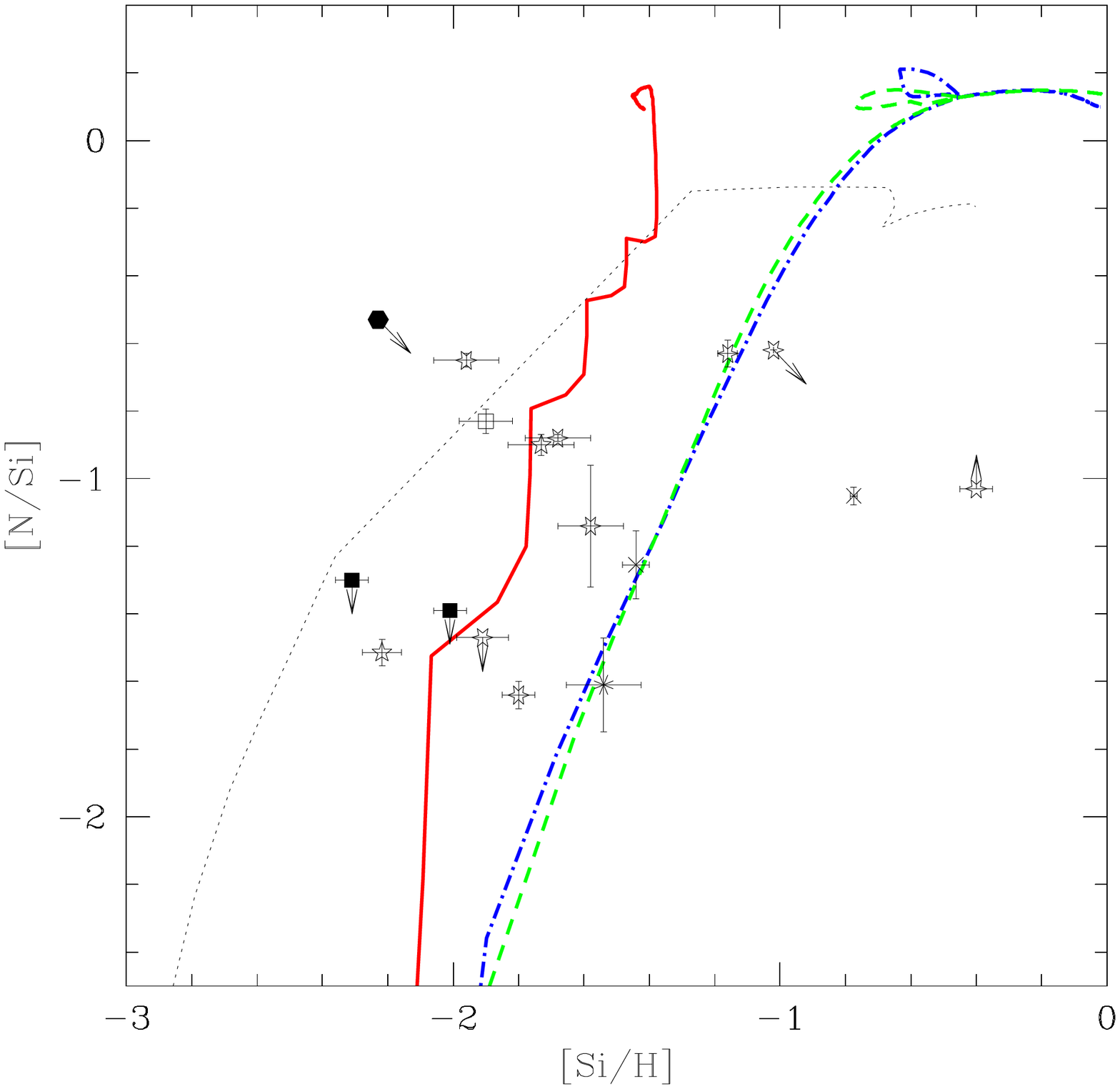}
                  \caption{[N/Si] vs [Si/H] as observed by various authors (see Calura et al. 2003 and references therein) and predicted 
by chemical evolution models. The models are the same as in figure 1.1. }
                   \label{sample-figure}
                \end{figure}

N is expected to be 
restored into the ISM after each $\alpha$ element, with a time delay depending
on the lifetimes of its stellar progenitors.
In particular, N is mainly produced in low and intermediate mass stars
which start dying after $\sim 3 \cdot 10^{7}$ yr since the beginning of 
star formation. The bulk of N production occurs after $\sim 250-300$ Myr
(Henry et al. 2000; Chiappini et al. 2003).
For this reason, the [N/$\alpha$] ratio is an important chemical evolution diagnostic 
and can provide useful information on the age of DLAs. 
In figure 1.4 we show the predicted [N/Si] vs [Si/H] and the values measured in DLAs 
by various authors (see Calura et al. 2003 and references therein). 
Unlike in the previous figures, here we do not show the curve for ellipticals since it 
would lie at [$\alpha$/H] values much higher than the ones observed in this sample of data.
There have been recent claims for a bimodal distribution of the [N/Si] vs 
[Si/H] values observed in DLAs, with a plateau located at [N/Si] $\sim -1.5$ 
including a very small set of objects, and the majority of the systems with 
[N/Si] $\ge -1.2$ (Prochaska, these proeedings; Molaro, these proceedings). 
Further observations are needed to investigate whether the data missing between 
the lower plateau and the bulk of the systems is real or due only to a small number 
of objects with available N measures.\\
 However, from figure 1.4 it is clear that spiral and irregular models can again fairly well reproduce the 
[N/$\alpha$] values observed in this sample of DLAs.

                \begin{figure}
                  \centering
                  \includegraphics[width=10cm,height=10cm,angle=0]{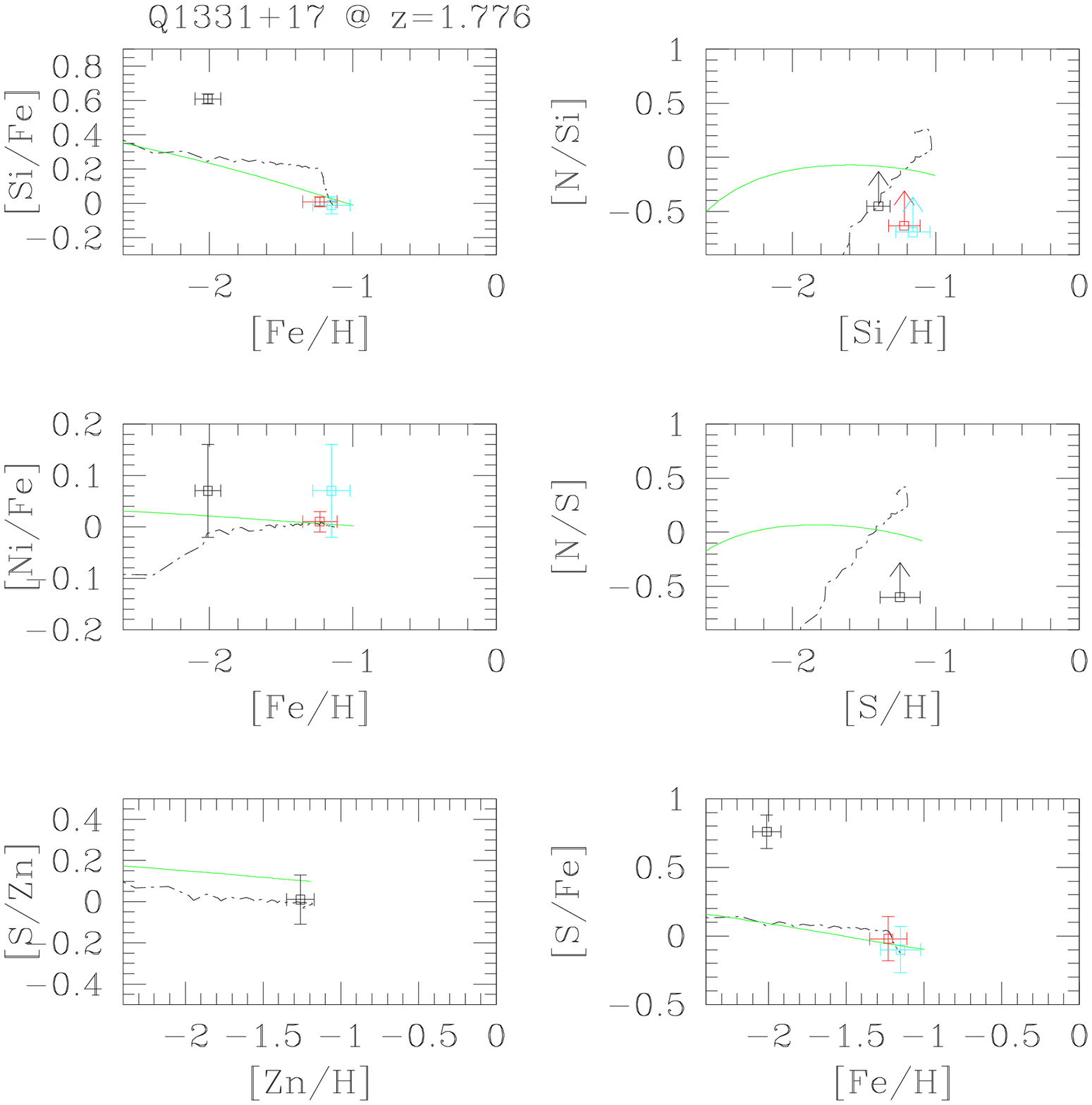}
                  \caption{Observed and predicted abundance ratios vs metallicity for the DLA towards Q1331+17. 
\emph{Black dash-dotted line:} model for a spiral galaxy at 16 kpc;
\emph{green solid line:} model for an irregular galaxy with continuous star formation and effiiency $\nu = 0.03 Gyr^{-1}$.
\emph{Black empty squares:} non dust-corrected measured abundance ratios;
\emph{red empty squares:} measured abundance ratios corrected for dust-depletion according to Vladilo (2002), model E00;
\emph{cyan empty squares:} measured abundance ratios corrected for dust-depletion according to Vladilo (2002), model E11.}
                  \label{sample-figure}
                \end{figure}
               
                  \begin{figure}
                    \centering
                  \includegraphics[width=10cm,height=10cm,angle=0]{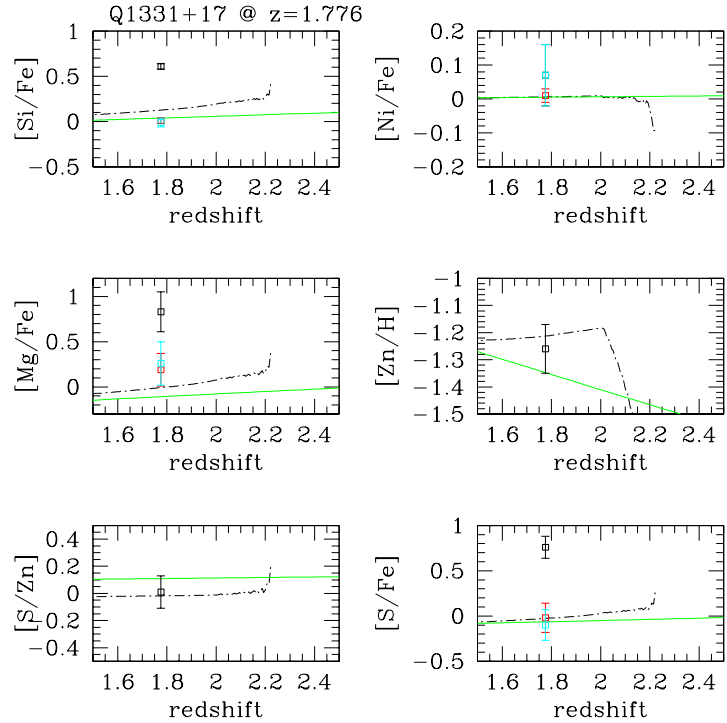}
                  \caption{Observed and predicted abundance ratios vs redshift for the DLA towards Q1331+17. The points are as in figure 1.5. 
The black dash-dotted line is for the outskirts of a spiral with formation redshift $z_{f}=2.8$, corresponding to an age of 1.44 Gyr
for a standard $\Lambda$CDM cosmology. The solid green line is for an irregular with low star formation efficiency ($\nu = 0.03 Gyr^{-1}$) 
and a formation redshift  $z_{f}>10$, corresponding to an age of $>3.4$ Gyr.}
		  \end{figure}

                \begin{figure}
                  \centering
                  \includegraphics[width=10cm,height=10cm,angle=0]{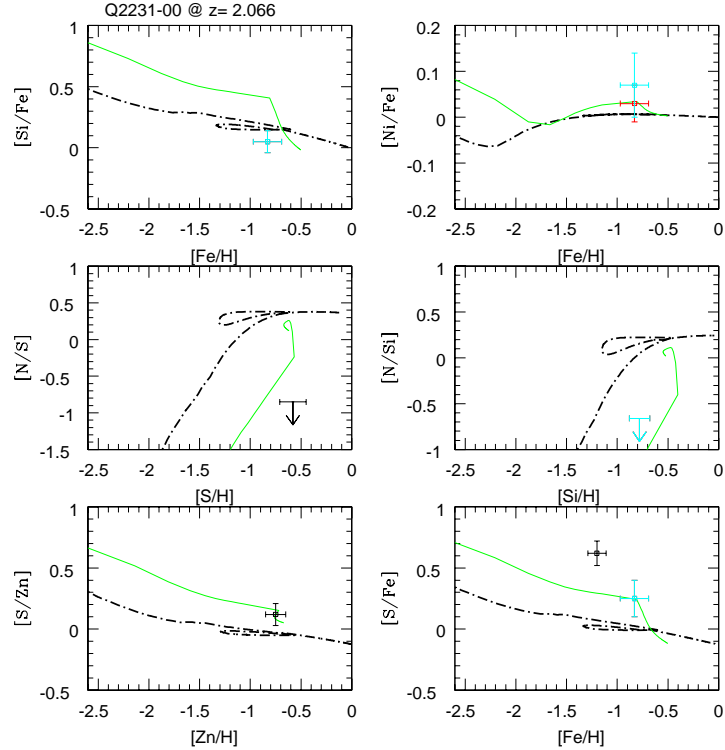}
                  \caption{Observed and predicted abundance ratios vs metallicity for the DLA towards Q2231-00. 
\emph{Black dash-dotted line:} model for a spiral galaxy at 4 kpc;
\emph{green solid line:} model for a dwarf galaxy with a starburst of duration $\Delta t = 0.1$ Gyr and efficiency $\nu = 4.2 Gyr^{-1}$.
\emph{Black empty squares:} non dust-corrected measured abundance ratios;
\emph{red empty squares:} measured abundance ratios corrected for dust-depletion according to Vladilo (2002), model E00;
\emph{cyan empty squares:} measured abundance ratios corrected for dust-depletion according to Vladilo (2002), model E11.}
                  \label{sample-figure}
                \end{figure}
               
                  \begin{figure}
                    \centering
                  \includegraphics[width=10cm,height=10cm,angle=0]{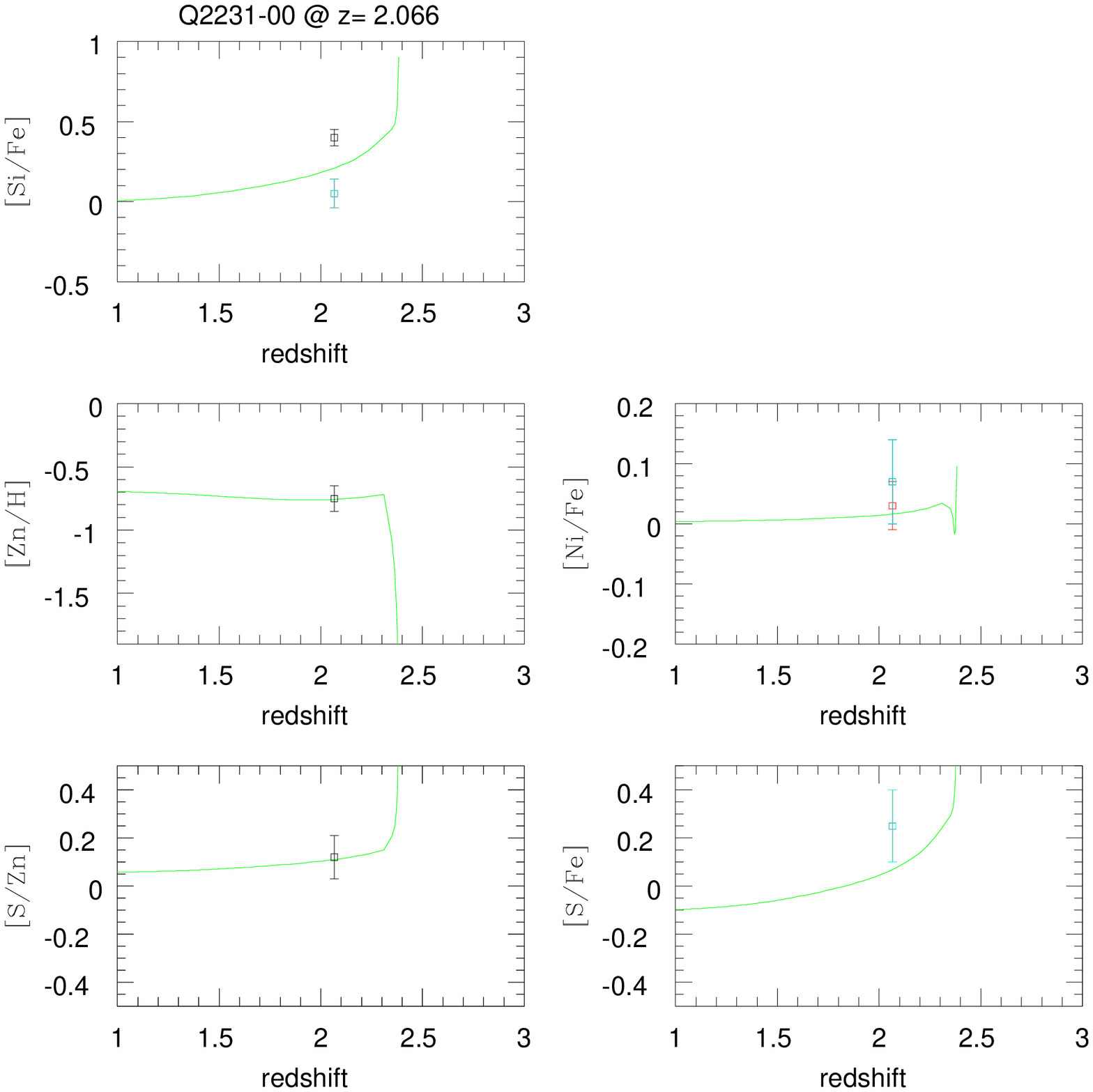}
                  \caption{Observed and predicted abundance ratios vs redshift for the DLA towards Q2331-00.
The points are as in figure 1.7. The green solid line is for a dwarf galaxy with a starburst of duration $\Delta t = 0.1$ Gyr, 
 efficiency $\nu = 4.2 Gyr^{-1}$ and formed at $z_{f}=2.8$, corresponding to an age of 0.9 Gyr.}
                \end{figure}

\subsection{Study of individual systems}
In previous studies, chemical evolution models were constructed in order to interpret 
the abundance patterns observed in DLAs as an ensemble, considering them as an evolutive sequence
(Matteucci, Molaro \& Vladilo 1997, Hou, Boissier \& Prantzos 2001).
By means of our chemical evolution models we have attempted a new approach: beside the study of the DLA population, 
we have focused on individual systems and tried to infer the nature of the associated galaxy and its age by studying the abundance 
patterns vs metallicity and redshift, respectively (Dessauges-Zavadsky et al., 2003, in preparation).  
Figure 1.5 shows the first results for the DLA towards QSO1331+17. 
The best fit is obtained with a model for the outskirts of a spiral galaxy or with an irregular galaxy with a 
continuous and low star formation. 
Figure 1.6 shows the results for the abundance ratios vs redshift plots: we can satisfactorily fit all the points by assuming a formation redshift of $z_{f}=2.8$ for the spiral (i.e. an age of 1.44 Gyr for the cosmology adopted here) or $z_{f} \sim10$ 
(i.e. an age of 3.4 Gyr) for the irregular.
In figure 1.7 we show the results for the DLA towards Q2231-00. In this case, the inner regions of a spiral model and a dwarf  galaxy with 
a starburst lasting $\Delta t = 0.1$ Gyr and efficiency $\nu = 4.2 Gyr^{-1}$ have provided the best fit to the data, though the starburst 
reproduces the observed pattern far more accurately than the disk model. In figure 1.8 we show the abundance pattern vs redshift for this DLA: 
here we have plotted only the starburst model prediction. The best fit is obtained assuming a formation redshift of $z_{f}=2.8$, 
corresponding to an age of 0.9 Gyr.

\section{Conclusions} 
In the present work, we have interpreted
the abundance patterns observed in DLAs by means of  detailed chemical
evolution models for galaxies of different morphological types,
i.e. ellipticals, spirals, irregulars/starbursts.  Our main
conclusions can be summarized as follows:\\
(i) owing to their very high
metallicity and [Si/Fe] ratios,  elliptical galaxies (and big spheroids in
general) represent unlikely candidates for DLA systems;\\
(ii) once the
observed abundances have been corrected by dust depletion, spiral and
irregular  galaxies are ideal DLA candidates; \\
(iii) we have shown that the study of
individual systems can provide very useful indications in DLA
studies:  the abundance ratios vs metallicity can help to constrain the
nature of the progenitor, whereas the abundance ratios vs redshift can
constrain its age. Here we have shown two cases: the abundance patterns observed in 
the DLA towards Q1331+17 are reproduced very well either by a model for the outskirts of a MW-like spiral or by an irregular galaxy with 
continuous and low star formation. The age of the system has turned out to be $1.44$ Gyr considering the spiral model, and $3.4$ Gyr in the case 
of the irregular one. The abundance ratios measured in the DLA towards Q2231-00 are reproduced by a starburst model with a burst of 
duration $0.1$ Gyr, 
a high SF efficiency ($4.2 Gyr^{-1}$) and formation redshift of $z_{f}=2.8$, corresponding to an age of $0.9$ Gyr.

 

\def\aa{{A\&A}}
\def\aas{{A\&AS}}
\def\aj{{AJ}}
\def\annrev{{ARA\&A}}
\def\apj{{ApJ}}
\def\apjs{{ApJS}}
\def\baas{{BAAS}}
\def\mnras{{MNRAS}}
\def\nat{{Nature}}
\def\pasp{{PASP}}


\end{document}